\documentclass[sigplan,screen,nonacm]{acmart}

\AtBeginDocument{%
  \providecommand\BibTeX{{%
    \normalfont B\kern-0.5em{\scshape i\kern-0.25em b}\kern-0.8em\TeX}}}

\usepackage{amsmath}

\usepackage{algpseudocode}

\usepackage{tikz}
\usetikzlibrary{arrows,positioning,shapes}

\usepackage{listings}
\newcommand*{\code}[1]{\lstinline{#1}}

\begin{document}

\title{Tracing OCaml Programs}

\author{Darius Foo}
\affiliation{%
  \institution{National University of Singapore}
  \streetaddress{}
  \city{}
  \state{}
  \country{}
  \postcode{}}
\email{dariusf@comp.nus.edu.sg}

\author{Wei-Ngan Chin}
\affiliation{%
  \institution{National University of Singapore}
  \streetaddress{}
  \city{}
  \state{}
  \country{}
  \postcode{}}
\email{chinwn@comp.nus.edu.sg}


\begin{abstract}
This presentation will cover a framework for application-level tracing of OCaml programs.
We outline a solution to the main technical challenge, which is being able to log typed values with lower overhead and maintenance burden than existing approaches.
We then demonstrate the tools we have built around this for visualizing and exploring executions.
\end{abstract}

\maketitle

\section{Overview}

The lack of good debugging tools is a frequently-mentioned pain point in the annual OCaml Users Survey~\cite{ocamlusers}.
Despite the variety of debugging tools and methods in the ecosystem (backtraces, \code{#trace}, ocamldebug, ocamli),
\emph{printf debugging} remains a serious option~\cite{whitington2020debugging,ocamldocsdebugging}, likely because it is highly \emph{accessible}~\cite{whitington2020debugging} -- it is available at full functionality under all circumstances.

To improve this situation,
we implement a framework for systematically capturing execution traces, and tools for filtering and viewing them in different ways to locate bugs.
This is done via source instrumentation, similar to classic tracers such as Hat~\cite{wallace2001multiple}, where a program is transformed so it logs its execution as it runs.
Tracing works well in production, pairs well with monitoring (which provides inputs to drive execution), and can be implemented as a \emph{ppx}~\cite{ppx} preprocessor, without requiring compiler or runtime changes.

What makes automatically tracing OCaml programs difficult?
One might imagine using a ppx to insert \emph{printf} statements into every function call.
The first problem with this is that OCaml does not yet have a mechanism for ad hoc polymorphism, so one cannot simply \code{show} values, and must supply a printer at each call site.
To generate such printers automatically, prior work such as \emph{typpx}~\cite{typpx}, \emph{typedppxlib}~\cite{typedppxlib}, and \emph{genprintlib}~\cite{genprintlib} invoke the typer (or read its outputs) during the ppx phase of compilation.
The downsides are that the generated code must be validated a second time by the typechecker after preprocessing, and these extensions depend heavily on compiler implementation details, often vendoring a copy of the typer's source code.

The next thing one might try is to log values in an untyped manner, e.g. with \code{Marshal}, and try to reconstruct the original values from the information available at runtime.
However, since OCaml verifies type safety at compile time, only a minimal amount of typed metadata is retained at runtime, e.g.~tags to distinguish variant constructors.
This leads to the same memory representations being used for different types of values. For example,
the value \code{(Some 1)} of \code{option} type is indistinguishable from the value \code{(Ok 1)} of \code{result} type in memory.
Any approach which seeks to ``show the programmer the system, not the machine''~\cite{halpern1965computer} must thus involve some compile-time component, to be aware of user-defined types.

The main novelty of our framework is a solution to this issue, using an alternative compilation pipeline which does not incur the cost of type-checking code multiple times.
The rest of the framework concerns tools for analyzing execution traces in various ways, motivated by the intuition that no single view is adequate for identifying every kind of bug~\cite{whitington2020debugging}.
The vision is that one can instrument their project with our framework automatically and readily record execution traces, which may be queried or replayed in tandem with the source code to assist in debugging, learning, and program comprehension.

In the rest of the presentation, we describe the design of our framework and the tradeoffs it makes.

\section{Design}

As mentioned earlier, the major difficulty in tracing OCaml programs is somehow being able to access type information to construct printers, without running the typer multiple times, obtrusive modifications to it, or vendoring its sources.

\begin{center}
\begin{figure}
\includegraphics[scale=0.35]{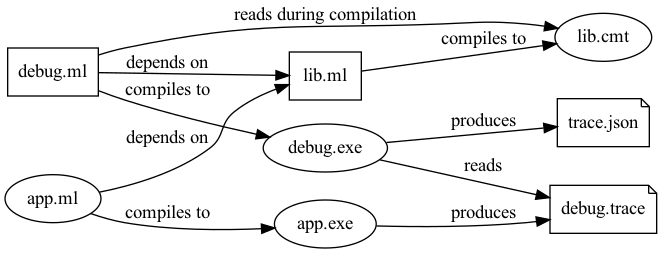}
\caption{Modified compilation pipeline}
\label{fig:compile}
\Description{Modified compilation pipeline}
\end{figure}
\end{center}

Designs for widely-used trace formats such as the Common Trace Format~\cite{ctf} offer a partial solution: to make traces as compact as possible, they separate trace content from metadata, so no space is wasted encoding structural elements such as delimiters. The metadata is separately passed to the decoder to enable it to read the trace.
Specializing this to OCaml, types are are not needed to output values, only to interpret them.

This idea serves as the basis for a compilation pipeline, shown in \autoref{fig:compile}.
\code{lib.ml} and \code{app.ml} represent a user \emph{library} (a module without an entrypoint) and \emph{executable} (a module with one); we assume \code{app.ml} is bug-free and only \code{lib.ml} is traced.
\code{debug.ml} is a separate executable users create to use our framework.
Rectangular nodes are processed by ppx.

\begin{enumerate}
  \item \code{lib.ml} is first compiled with a ppx to instrument functions, serializing values using \code{Marshal}.
  \item \code{debug.ml} is compiled with a \emph{second} ppx which reads the \code{cmt} files of \code{lib.ml}, generating code which is aware of how to interpret marshalled values.
  \item \code{app.exe} is run to produce a trace.
  \item \code{debug.exe} is run to read the trace and convert it into a form that can be more easily queried by downstream tools.
\end{enumerate}

This essentially introduces an explicit form of staging to the build, allowing the compilation of a module to depend on the \emph{types} of another.
As \code{cmt} files are used, the typer is only run once on the latter, and the former does not incur any compilation overhead if the trace is never read.
The approach also inherits the benefits of compact trace formats, incurring less \emph{runtime} overhead -- a reason to continue doing this even when some form of ad hoc polymorphism arrives in OCaml
The maintenance burden is low, as outside the public APIs of the compiler, only \code{Cmt_format} is depended on -- the typer does not need to be vendored.

The tradeoffs are that it is the \emph{build} that is now non-standard, and it does not have the same expressiveness as an approach like \emph{typedppxlib}~\cite{typedppxlib}, which can use types in arbitrary ways at compile time.
Notably, this explicit staging does not work when type-dependent values must be read by other parts of \code{lib.ml}.
Nevertheless, this compilation pipeline may be useful beyond the current work for things like type-safe deserialization in RPC frameworks such as \emph{protobuf}~\cite{protobuf}.









\section{Visualization}

Given a way to output arbitrary typed values in OCaml programs, we now consider how to present the data.
We could output a simple sequence of events, but as we currently log function arguments and return values, as well as \code{match} discriminees, we instead arrange them into trees of calls and use that as the common format for downstream tools.

\begin{center}
\begin{figure}
\includegraphics[scale=0.35]{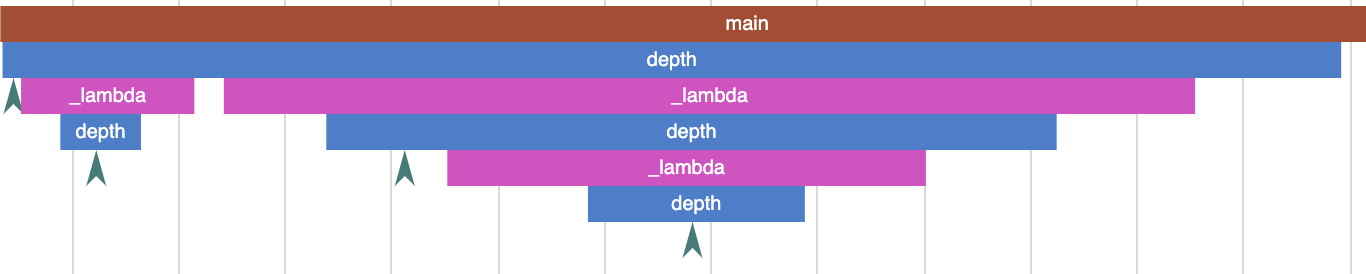}
\caption{Flame graph rendering of \code{depth}}
\vspace{-1em}
\label{fig:depth}
\Description{Flame graph rendering of depth}
\end{figure}
\end{center}

We have prototyped a number of user interfaces for analyzing traces.
The simplest ones are batch CLI tools, for printing the call tree, or finding all instances of specific calls (like \code{hat-observe}~\cite{hat}).
Others produce other trace formats, such as Chrome's \emph{Trace Event Format}~\cite{chrometrace},
which can be given to
Magic Trace~\cite{magictrace} to render as a flame graph~\cite{flamegraphs}.
Magic Trace also provides the ability to query traces using SQL.

\begin{center}
\begin{figure}
\begin{lstlisting}[language=caml]
let rec depth t =
  match t with
  | Leaf _ -> 0
  | Node sub -> List.fold_right
      (fun c t -> max (depth c) t) sub 0 + 1
\end{lstlisting}
\caption{A simple recursive function}
\vspace{-1em}
\label{fig:depthsrc}
\Description{A simple recursive function}
\end{figure}
\end{center}

A simple recursive function for computing the depth of a rose tree (\autoref{fig:depthsrc}) is rendered as shown in \autoref{fig:depth}.
In the diagram, time moves to the right and the stack grows downwards.
The chevrons represent \code{match}es, capturing the fact that the first thing that occurs in the call to \code{depth} is the \code{match}, followed by a call to the higher-order function by \code{fold_right} (which itself isn't traced, as it is not in \code{lib.ml}).
This is followed by a recursive call to depth, which \code{match}es before returning.
After that there is another call to the higher-order function from \code{fold_right}, and so on.
Clicking on each bar shows the argument and return values in OCaml-like syntax.
Argument values flow downwards as the stack grows, and return values flow upward.

We have implemented another visualizer using VSCode's Debug Adapter Protocol, which allows users to navigate traces as if they were in an interactive debugger, backwards and forwards, as they are unconstrained by actual execution.
This is a combination of the classic tools \code{hat-trail} and \code{hat-explore}~\cite{hat}.

These tools are all complementary, supporting different debugging workflows.
For example, exceptions are difficult to render in a flame graph view, as it is difficult to display stack unwinding in a purely additive way: the frame of the handler could be any of those up the stack in the flame graph.
Understanding exceptions in the user interface of an interactive debugger is comparatively straightforward, as users can see the jump in control flow occur, and navigate backwards and forwards to confirm the behaviour as they wish.
Another example where reverse execution helps is exploring computations backwards, starting from an error and going back to find context about its cause.
The flame graph view and CLI tools could be used to locate a particular buggy call, which could serve as a starting point for interactive analysis.

\section{Future Work}

The goal of this project is to make understanding what OCaml code does easy and widely-applicable.
In addition to debugging, we hope
the multiple interactive views of traces
can help the
exploration of new codebases and teaching.

The main limitation is that
our framework requires the project being traced to be recompiled,
in order to make use of type information, so it requires source code.
Library code also isn't traced for this reason, so it stops being useful for debugging once control flow leaves user code.
Because of that it is best used in projects which do not have many external dependencies; a nice application area is programming language tools.

We hope to investigate Magic Trace (and its use of Intel PT) to see whether it can be combined with what we have done to lift these restrictions.

\bibliographystyle{ACM-Reference-Format}
\bibliography{sample-base}










\end{document}